\documentstyle[12pt,titlepage]{article}
\newcommand{\ce}[1]{\centerline{\bf{#1}} \vskip .2in}
\newcommand{\cen}[1]{\centerline{#1}}

\begin{document}

\newcommand{\su}{\hspace*{.1in}}
\newcommand{\sdue}{\hspace*{.2in}}
\newcommand{\stre}{\hspace*{.3in}}
\newcommand{\squa}{\hspace*{.4in}}
\newcommand{\scin}{\hspace*{.5in}}
\newcommand{\ssei}{\hspace*{.6in}}
\newcommand{\para}{\par}
\newcommand{\be}{\begin{equation}}
\newcommand{\en}{\end{equation}}
\newcommand{\sot}{\hspace*{.8in}}
\newcommand{\lop}{\stackrel{<}{\sim}}
\newcommand{\gop}{\stackrel{>}{\sim}}
\def\gte{\lower 0.5ex\hbox{${}\buildrel>\over\sim{}$}}
\def\lte{\lower 0.5ex\hbox{${}\buildrel<\over\sim{}$}}
\def\loe{\lower 0.6ex\hbox{${}\stackrel{<}{\sim}{}$}}
\def\goe{\lower 0.6ex\hbox{${}\stackrel{>}{\sim}{}$}}
\newcommand{\ggg}{$\gamma$}
\newcommand{\eee}{$e^{\pm}$}
\newcommand{\lap}{$L_{38}^{-1/3}$}
\newcommand{\ergs}{\rm \su  erg \su s^{-1}}
\newcommand{\etal}{ {\it et al.}}
%
\def\jref#1 #2 #3 #4 {{\par\noindent \hangindent=3em \hangafter=1
      \advance \rightskip by 0em #1, {\it#2}, {\bf#3}, #4.\par}}
\def\rref#1{{\par\noindent \hangindent=3em \hangafter=1
      \advance \rightskip by 0em #1.\par}}
\newcommand{\porb}{ P_{orb} } 
\newcommand{\Po}{$ P_{orb} \su$}
\newcommand{\pdot}{$ \dot{P}_{orb} \,$}
\newcommand{\pot}{$ \dot{P}_{orb} / P_{orb} \su $}
\newcommand{\s}{ \\ [.15in] }
\newcommand{\mm}{$ \dot{m}$ }
\newcommand{\mdot}{$ |\dot{m}|_{rad}$ }
\newcommand{\myr}{ \su M_{\odot} \su \rm yr^{-1}}
\newcommand{\msol}{\, M_{\odot}}
\newcommand{\ppp}{ \dot{P}_{-20} }
\newcommand{\ci}[1]{\cite{#1}}
\newcommand{\bb}[1]{\bibitem{#1}}
\newcommand{\ch}[1]{\vskip .3in \noindent {\bf #1} \para}
\newcommand{\cms}{ \rm \, cm^{-2} \, s^{-1} }
\newcommand{\nn}{\noindent}
\newcommand{\asca}{{\it ASCA} }
\newcommand{\syn}{synchrotron }

\nn

\nn


\newcommand{\gro}{GRO~J1655-40 }
\newcommand{\grop}{GRO~J1655-40}
\newcommand{\ergsp}{\rm \;  erg \; s^{-1}}
\newcommand{\fs}{.''}

\cen{\large \bf The Dual Nature of Hard X-Ray Outbursts from }
\cen{\large \bf the Superluminal X-Ray   Transient Source \gro}

\vskip .3in
\cen{\large M. Tavani\footnote{
Columbia Astrophysics Laboratory, Columbia University, New York, NY 10027},
A. Fruchter\footnote{
Space Telescope Science Institute, 3700 San Martin Drive, Baltimore,
MD 21218},
 S.N. Zhang\footnote{
Universities Space Research Association, NASA/MSFC, Huntsville, AL 35812},
  B.A.  Harmon\footnote{
NASA Marshall Space Flight Center, Huntsville, AL 35812}}
\cen{\large R.N. Hjellming\footnote{
National Radio Astronomy Observatory, Socorro, NM 87801-0379},
M.P. Rupen$^5$, C.  Bailyn\footnote{
Department of Astronomy, Yale University, PO Box 208101, New Haven,
CT 06520}
 \& M. Livio$^2$} 

\vskip .4in

\baselineskip 19pt

\ce{Abstract}

We report the results of multiwavelength  observations
of the superluminal X-ray transient  \gro during  and following
the  prominent hard X-ray outburst  of March-April 1995.
\gro was continuously monitored by 
 BATSE on board
 CGRO,
and repeatedly observed in the radio and optical bands from the ground.
About a month after the onset of the hard X-ray outburst,
\gro was observed twice by HST on April 25 and 27 1995,
with the aim of detecting faint optical
emission from ejected plasmoids.
Despite the similarity of the hard X-ray emission 
in April 1995
with previous events in 1994, 
no radio or optical emission from \gro related to plasmoids 
was detected.
We conclude that
\gro  is subject to a 
complex behavior showing:
radio-loud 
hard X-ray outbursts with strong  radio  emission (of flux $f_r \goe 100$~mJy) 
both from a `core' source and from propagating  plasmoids 
(as those in 1994),
and 
radio-quiet hard X-ray outbursts with no detectable
 radio emission and plasmoid activity ($f_r \loe 0.5$~mJy)
(as those in 1995).
Our results can constrain models of particle acceleration and radiation of
relativistic plasmoids.

\vskip .7in
\cen{Submitted to the {\it Astrophysical Journal Letters}: July 31, 1996}
\cen{Accepted: October 2, 1996}

\vskip .1in
\cen{{\it Subject Headings:} acceleration of particles, instabilities, MHD}
\cen{radiation mechanisms: non-thermal, X-rays: stars, radio continuum: stars}

\newpage
\ce{Introduction}

Superluminal  X-ray  transients 
form a newly discovered class of  Galactic sources
showing strong  hard  X-ray outbursts
and sporadic
 relativistic ejection of radio
emitting plasmoids  in  a double-jet geometry.
 Qualitatively, the jet geometry of superluminal X-ray transients resembles 
that of 
AGNs and  SS~433, suggesting similar
conditions of accretion onto their respective
 compact objects.
Two clearly established  superluminal transients
are currently known in the  Galaxy, GRS~1915+105 (Mirabel \& Rodriguez 1994),  
and \gro (Harmon \etal\ 1995, hereafter H95; Hjellming \& Rupen 1995,
hereafter HR95; Tingay \etal\ 1995, hereafter T95).
 
GRO~J1655-40 (Nova Scorpii 1994)
 was discovered at the end of July 1994 by 
the Burst and Transient Source Experiment (BATSE) on board 
the {\it Compton} Gamma-Ray Observatory (CGRO)
 near the Galactic plane (Zhang \etal\ 1994).
The distance derived from the jet kinematics of 3.5~kpc (HR95)
is supported by the measurements of the 21-cm H~I 
(McKay \& Kesteven 1994, T95),
X-ray absorption (Inoue \etal\ 1994),
and optical absorption (Horne \etal\ 1996).

  GRO~J1655-40
exhibited  three prominent hard X-ray outbursts during 1994
(H95)
 (approximate starting dates: 1994 July 31, September 6, November 10).
In all of
these outbursts, \gro showed remarkable radio flaring in the GHz range
during or immediately following episodes of intense hard X-ray emission 
(H95, T95, HR95).
For these outbursts,
superluminal  motion of radio plasmoids 
was repeatedly observed (HR95).
 From  VLA and VLBA maps obtained during the period
 of August-November 1994, it was possible to deduce a time variable
 geometry of plasmoid ejection in oppositely-directed jets (HR95) 
resembling
the precessing jets of
 of SS~433 (e.g., Margon 1984). 
 A  kinematic model for the major  plasmoid ejections  and propagation 
in 1994 gives
  a plasmoid velocity $\beta_p = v/c 
= 0.92 \pm 0.02$,
 and an angle between the
radio jet axis and
 the line of sight direction, $i \sim 85^{\circ}$ (HR95).

Optical photometry 
obtained after the first
hard X-ray outburst
 in August 1994, revealed a brightened optical
counterpart with a  visual magnitude  in the range $V\sim 14-16$,
 and a change from the quiescent magnitude $V_q \sim 17.3$
(Bailyn \etal\ 1995, hereafter B95a).
The red color of the optical counterpart with
$E(B-V) = +1.3\pm 0.2$ as recently determined 
by deep 200~nm HST observations in May 1995 (Horne, Haswell \etal,
1996)
 indicates  significant absorption
in the direction of GRO~J1655-40   (see also B95a).  
 Spectroscopic CTIO  observations of \gro carried out 
 in early  May~1995 revealed Doppler-shifted high-excitation emission lines
 superposed on an F- or early-G-type stellar
 absorption spectrum (Bailyn \etal\ 1995b, hereafter B95b).
  If due to a companion star, the line velocity profile
 indicates an orbital period $P_{opt} = 2.62 \pm 0.03$~days
 and   inferred mass function 
 $f(M_2) = 3.16 \pm 0.15 \, M_{\odot}$ (B95b).
 A large mass function together with the constraints on the mass of the
companion star 
 favor  an interpretation of \gro  in terms of a black hole system 
 of primary mass $m \goe 3 \, M_{\odot}$
(B95b, see also Zhang \etal\ 1996).

Motivated by the apparent 
repetitive behavior of \gro during the August 1994-January 1995
outbursts (hard X-ray  emission, strong radio emission,  plasmoid propagation
with  associated radio emission lasting for several weeks),
we planned a series of multiwavelength observations. 
Previous activity  of \gro in 1994
showing a recurrent  2-3 week delay between the hard X-ray emission onset and
peak radio emission and plasmoid propagation  (H95, HR95)
allowed
us
 to plan an observational  campaign extending in time for about 1 month.

The opportunity for activating the planned campaign was offered in
March 1995, when BATSE detected a strong
hard X-ray outburst of  \gro 
 with characteristics very similar to the
previous ones (Wilson \etal\ 1995).
 The subsequent weeks of activity of the source,
coincided with 
previously scheduled ground based optical
and radio observations. Furthermore, a target of opportunity HST
observation was granted  and  scheduled within $\sim 1$~month of the 
hard X-ray outburst onset.

\vskip .2in
\ce{Observations}

\nn
{\it BATSE}

BATSE
can continuously monitor hard X-ray sources in the 20 keV$-$2 MeV band
by the Earth
occultation technique (Harmon \etal\ 1992).
 BATSE effectively monitored hard X-ray emission of  \gro  since its
 discovery in late July 1994 (H95).
Figure~1 shows the BATSE (one-day averaged)
 light curve of hard X-ray emission in the 
20-100~keV band extending up to August~1995.
A prominent hard X-ray outburst was detected during the period March-April 1995
($\sim$~MJD~9770-9820) with peak flux corresponding to $\sim$1.3~Crab units
around March 24, 1995 (MJD~9800). Both the rise and decay parts of the 
light curve have similar characteristics and the whole hard X-ray outburst has
a `triangular' shape.

 \vskip .1in

\nn
{\it Radio observations}

Radio observations of \gro were carried out at the Very Large Array (VLA)
throughout the whole period of the continuous BATSE monitoring.
The 1995 observations were made at 4.9 and 8.4 GHz, and in all cases
no radio emission was detected with an upper limit of 0.5 mJy.
Figure~1 shows the complete (logarithmic)  1994-1995 VLA radio light curve of 
\gro at 1.49 GHz for 1994 and the 4.9/8.4 GHz upper limits in 1995, plotted 
on the same temporal scale as the BATSE data.
The prominent 1994 radio flares of \gro in coincidence with the
hard X-ray outbursts are clearly evident (Harmon \etal\ 1995, HR95).
The last VLA radio detection of \gro occurred  near mid-December 1994
in coincidence with the final decay of the November-December 1994 hard X-ray
outburst. Since then, the radio source was not detected again until
the recent flare\footnote{
   The 1996 X-ray flare was detected by the
Rossi X-ray Timing Explorer's All Sky Monitor (ASM)
(Remillard \etal\ 1996)
on 1996 April 25.  Horne \etal\ (1996) reported the optical/ultraviolet
brightening of \gro on May 10.  VLA upper limits for the detection
of radio emission from this object were obtained, after the onset of
the X-ray flare on, May 10, 12, and 20.  With the Hunstead
and Campbell-Wilson (1996) detection of 55 mJy at 843 MHz on May 28 
and the Hjellming and Rupen (1996) detection of 19 mJy at 4.9 GHz
on May 29, the onset of the radio flare must have been between 24 and
32 days after the beginning of the X-ray outburst detected by
the ASM.
A detailed account of the 1996 emission event 
will be presented elsewhere.}
 in 1996 (Hunstead \& Campbell-Wilson 1996, Hjellming \& Rupen 1996).

The diversity of the radio/hard X-ray correlation between the 1994 and
1995 behavior of the source is evident from Figure~1.
Contrary to previous activity in 1994, the March-April 1995
(and  mid-August 1995) hard X-ray outbursts were not followed by detectable
radio flaring.  In particular, there was no detectable radio emission
at the time of the HST observations. as indicated by the vertical lines
superposed on the radio light curve in Figure~1.
What may be a third class of behavior was seen for the 1996 flare
event.

\vskip .1in
\nn 
{\it HST observations}

Following the March 1994 hard X-ray ourburst of \grop,
 HST observations aimed at detecting optical emission from 
plasmoids
were proposed.  
The relative proximity  of the 
radio
plasmoids to  the central source 
after 1-2 weeks of propagation as observed in 1994
  ($0.'' 3 - 0.''4$,  HR95, T95), the brightness of the
central source (B95a,b), and the faintness of the estimated optical
emission of plasmoids 
all require the superior imaging capability of the refurbished HST.

 Target of opportunity 
HST-WFPC2  observations of \gro were carried out on April 25 and 27, 1995.
A red  filter (F675W) 
was chosen to maximize the possibility of detection given  
the extinction along the direction of 
the source.
  Two visits, separated by approximately 48 hours, 
were planned, in order
to determine
the 
velocity of any ejecta that might be detected.  
On each visit, we obtained four 40~s exposures in each of four positions,
using the PC camera of the HST WFPC2.  The four positions were placed on
a parallelogram, whose vertices were offset by non-integral pixel
shifts.   These
positions fully sampled the instrumental point spread function (PSF).  
In addition, on the second visit the telescope was rolled by approximately
25 degrees with respect to  the orientation of the first visit. 
Rolling the telescope causes any non-axisymmetric features of the PSF
to rotate with respect to the sky, and thus 
it simplifies the task of 
distinguishing between faint PSF artifacts and astrophysical emission.

The four images at each dither position
were then cleaned of cosmic rays and averaged.  The
four resulting images from each visit were finally  combined 
by shifting and adding.  Figure 2
 shows the image obtained during the first 
 April 25 1995 visit
(source position:
 R.A. = 16h 54m 00.137s, DEC =   --39$^{\circ}\, 50' \, 44.90" \pm
 0.20"$  (equinox J2000).  
The central  source has a Cousins  average apparent magnitude
(band pass 0.57-0.72~$\mu$)
of $R_c = 16.22 \pm 0.03$ in the first visit,
 and $R_c = 15.97 \pm 0.03$ for the  April 27
visit.
 No sign  of additional optical emission, other than that of the central
point source, was  immediately obvious in either image.  As a further
check, the sub-sampled images were scaled and subtracted
from one another.  The  difference image was then 
deconvolved 
with the PSF and the residuals measured.   No emission of flux  
higher
than 0.03 that
of the star was found within $0\fs2$ of the stellar position.
 The limit within $0\fs5$ is $ \sim 0.002$ and beyond $0\fs7$ the limit is
$\sim 0.0005$ that of the stellar flux.
The apparent $R_c$ magnitude upper limits 
to plasmoid optical emission are in the range
$20 \loe R_c \loe 24$.
The apparent $R_c$ magnitudes  need to be corrected for
extinction, and the corrected upper limits
are $R'_c = R_c - A_{R_c}$, with $A_{R_c}=3.25$ (assuming $E(B-V) = 1.3$).
Table~1 summarizes the relevant HST upper limits to optical emission
of plasmoids from \gro in late April~1995. 

\vskip .2in
 
\ce{Discussion}

Our observations 
can
constrain models of particle acceleration and radiation
of the  relativistic plasmoids of \grop.
The particle composition and the energy spectrum of particles in the plasmoids
are  unknown. The only established fact is the radio emission in the
$10^9-10^{10}$~Hz frequency range as observed in 1994 (HR95, T95).
This radio emission
implies a relativistic Lorentz factor for the radiating
electrons/positrons at the distance 
$r'$,  $\gamma_{e,r} \sim 7.5 \, \nu_9^{1/2} \, B(r')^{-1/2}$,
where $\nu_9$ is the photon frequency  in units of $10^9$~Hz and 
$B(r')$ the local magnetic field  in Gauss 
(e.g., Pacholczyk 1969).
A single distribution of particle energies $N(\gamma_e) \propto
\gamma_e^{-\delta}$ with $\gamma_{min} \leq \gamma_e \leq \gamma_{max}$
may produce (self-Compton) synchrotron  radiation  in the optical band
for a variety of conditions including adiabatic expansion and radiation
(shock re-energization along the jet may produce an effective $N(\gamma_e)$
more complex than a single power-law). If the Lorentz factor of the optically
radiating particles $\gamma_o$ satisfies the relation
$\gamma_{min} \leq \gamma_{e,r} \ll \gamma_o \leq \gamma_{max}$,
with $\gamma_o \sim 300 \, \gamma_{e,r}$,
synchrotron  radiation may occur from the radio to optical band
as observed in several extragalactic jet sources (e.g., M87,
Biretta \etal\ 1993). We call this model of emission
model (1).
Alternately, the low-energy 
 cutoff of $N(\gamma_e)$ may occasionally
be shifted at higher energies, with $\gamma_{min} > \gamma_{e,r}$,
or \syn self-absorption can occur.
In this case, a `radio-invisible' plasmoid can be produced, 
and depending on the ratio of $\gamma_{min}/\gamma_o$ and on 
the self-absorption frequency,  optical emission
may still be detectable. We call this model of emission model (2).
Deep optical observations of propagating plasmoids of
Galactic superluminal transients  can in 
principle constrain $N(\gamma_e)$ and its range.

A synchrotron  model of emission assumes a plasmoid  containing
relativistic electrons (and positrons, if present) plus
 baryons being ejected
with a  bulk velocity $\beta_p$ and subject to radiative and
adiabatic losses.
The radiating particles are likely  to be impulsively
accelerated by a shock or by a MHD  acceleration mechanism in a way
similar to
that proposed for
 jets of  extragalactic sources and young stellar objects
  (e.g., Blandford  1993;  K\"onigl,  1989).
After an initial phase dominated by optically thick emission,
the \syn emission model
leads in general to a broken power law emission extending
from radio to higher frequencies.
The observed  slope of the 1994 radio  emissivity
(in the frequency range $10^9-10^{10}$~Hz)  from the
plasmoids of \gro is $\alpha_o \sim 0.4-0.6$ (HR95).
The spectral slope may change due to synchrotron
losses to $\alpha_0 + 1/2$ at a `break frequency'
$\nu_b$ which 
depends on the magnitude of the
magnetic field convected by the plasmoid along the jet
(e.g., K\"onigl 1981).
The observed\footnote{
The true luminosity $L_t$ for  jet-related emission towards the
Earth is related to the apparent luminosity
by $L_t = (1 - \xi) \, L_r/b$, where
$\xi$ is the fraction of radio luminosity produced by the core
source and by the counter-jet, and $b = [(1 - \beta_p^2)^{1/2} \,
(1 - \cos i)^{-1} ]^{3 + \alpha_o}$.}  
radio luminosity
near $10^9$~Hz
 of an individual plasmoid ejected from \gro
is $L_r \sim 10^{31} \, f_{J} \, \ergs$ where $f_{J}$ is
the radio flux (in Janskys).
Observations of radio flares  of \gro show that $f_J \sim 1$
for the early part of major plasmoid ejections in 1994 (HR95).
One can then calculate the expected optical luminosity of
\gro plasmoids  
for a variety of acceleration and emission  characteristics.
For simplicity, we consider here only general
properties of the \syn emission model.
In principle, optical emission from plasmoids can be also produced 
by  a synchro-Compton  mechanism or by internal or external
shocks generating continuum and line emission.
A systematic discussion of optical emission mechanisms of
Galactic transient plasmoids   will appear elsewhere.

The  radio spectra
of the 1994 plasmoids 
suggests an energy index $\delta \sim 2$.
 The  optical
luminosity in the $R_c$ band  $L_o(R_c)$
from the \syn mechanism can therefore be estimated as 
  $L_o/L_r  \goe 200$, for $\nu_b > 10^{14}$~Hz. 
If  plasmoids  were produced as in the 1994 flares
with $f_J \sim 1$,
optical emission in the $R_c$-band would have been detectable by our HST
observations for an estimated observable
 luminosity $L_o \sim  3\cdot 10^{33} \ergs$.
For lower radio emission, a \syn model with $\nu_b > 10^{14}$~Hz and
$\delta=2$ 
can be tested by 2-visit HST observations for plasmoids emitting 
radio emission of flux  $f_J \goe  0.1$ within
$0.''5$ of the central source, and for
$f_J \goe  0.01$ within an angular distance of $1''$
(see Table 1).
This optical emission
is expected to be  simultaneous with the appearance
of a peaking radio flare from the optically thin jet.

The lack of detected radio plasmoids during our 2-visit
HST observations
does not allow the testing\footnote{ 
Table~1 shows that
an improvement in flux sensitivity by two orders 
in magnitude 
would be necessary to constrain model (1) for $f_J \sim 0.0005$
corresponding to the VLA upper limit of 0.5~mJy.}
of model (1).  However,
model (2) can be constrained by our observations with the assumption
of a plasmoid fractional energy available 
to radiation  similar to the 1994 episodes,
for $\gamma_{min} \sim \gamma_o$ and $L_o \sim  3\cdot 10^{33} \ergs$.
 We find  that the total
energy of the possibly radio-invisible plasmoids of
\gro in March-April 1995 is constrained to be less by approximately
two orders of magnitude  compared to the first August 1994 event.

\vskip .2in
\ce{Conclusions}

The behavior
of \gro
can be remarkably dissimilar  for different hard X-ray  outbursts.
 The lack of detectable radio and plasmoid emission in April 1995
 establishes, for the first time in a
Galactic superluminal transient, the 
`dual' nature 
which is probably
of the accretion process
producing the  hard X-ray outbursts.
We can therefore establish the existence of   two kinds of 
hard X-ray outbursts for \grop:

\begin{itemize}
\item
radio-loud
 hard X-ray outbursts, as those characterized by major radio flares
of the central source and associated plasmoid propagation detected in 1994.
The radio luminosity (core plus plasmoid) was observed to be in the range\\
$10^{29} \ergs \loe L_r \loe 10^{32} \ergs$ for a source at a
3.5~kpc distance.
The ratio $\eta_R $ of  radio to hard X-ray  luminosities of 
radio-loud events
 is therefore $10^{-8}/b \loe \eta_R  \loe 10^{-5}/b$, with 
$b$ the  beaming factor defined above
($b \sim 0.05$ for the jet directed toward the Earth);

\item
radio-quiet
 hard X-ray outbursts,
as those observed in 1995 and  characterized by the absence of detectable
radio emission with a ratio $\eta_Q $ of  radio to hard X-ray  luminosities
satisfying the relation $\eta_Q \loe 10^{-9}/b$.
No optical emission down to $R'_c \sim 20$ is detected
at $\sim 0."5$ from the central source,
making 
the existence of `radio-invisible' jets
implausible.
\end{itemize}

The upper limit on radio emission for 
the radio-quiet hard X-ray outbursts
of \gro  is about three
orders of magnitude smaller than the 
brightest
emission  of
radio-loud outbursts.
It is then clear that the accretion processes leading to 
strong  hard X-ray  emission are not 
uniquely  related to  relativistic plasmoid  energization
and outward propagation.
Hard X-ray emission 
was always detected 
preceding the 1994
major radio flares of \gro  (H95, HR95).
However, similar hard X-ray outbursts of \gro can have a very different
behavior
in the radio band, as shown by the 1995 activity.
It is interesting to note that a somewhat similar behavior was found in the
broad-line radio galaxy 3C~390.3.
In a recent paper, Eracleous, Halpern \& Livio (1996) have shown that
the ejection of radio blobs in this object does not appear to
be correlated with any fluctuation of the X-ray flux in the 1-10~keV band.
This could mean that the ejection of plasmoids responds more to
instabilities in
the plasma outflow or corona rather than to changes in the accretion rate.

We also note that  possibly elongated near-IR emission was
detected in July 1995 
from the superluminal jet source GRS~1915+105 (Sams, Eckart
\& Sunyaev 1996). If due to jets, this near-IR emission might indicate
the existence of an energetic population of particles
in the plasmoids.
However,
 simultaneous radio observations of GRS~1915+105 (Foster \etal\ 1996)
show a low-intensity
flux (at 2-8~GHz) well below a simple power-law extrapolation from IR to
radio bands. We deduce that the elongated IR emission from 
GRS~1915+105 was due to either
synchrotron jet emission  strongly absorbed in the radio band,
or  interaction with the environment of radio-invisible jets.
Our April 1995 observations  exclude 
both these possibilities for \grop.

The exceptional nature of \gro stimulated an unprecedented 
coverage
of its X-ray and radio emission during an extended period of time.
Future observations will determine
whether the dual outbursting behavior of \gro is the norm or the exception
among the superluminal transients
 and black hole candidates.

 \vskip .2in

We thank the Space Telescope Science Institute management and Director for
the prompt response in granting and scheduling the
target of opportunity observation of \grop.
Research partially supported by NASA grants HST-GO-6248,
NAG~5-2729, and NAGW~2678.

\newpage
\vskip .2in 
\noindent
{\bf References}

\jref{Bailyn, C. \etal,  1995a} {\it Nature} {\bf 374} {701 (B95a)}

\jref{Bailyn C., \etal, 1995b} Nature 378 {157 (B95b)}
 
\rref{Biretta, J.A., 1993, in
{\it Astrophysical Jets}, eds. D. Burgarella, M. Livio \& C. O'Dea
(Cambridge University Press), p. 263}

\rref{Blandford, R.D.,  1993, in
{\it Astrophysical Jets}, eds. D. Burgarella, M. Livio \& C. O'Dea
(Cambridge University Press), p. 15}

\jref{Eracleous, M., Halpern, J.P. \& Livio, M., 1996} ApJ 459 89

\rref{Foster, R.S., Waltman, E.B., Tavani, M.,
Harmon, B.A. \& Zhang, S.N., 1996,
{\it Astrophys. J. Letters}, {\bf 467}, L81}

\rref{Harmon, B.A., \etal, 1992, in Compton Observatory Science
Workshop (Proceedings), eds. C. R. Shrader, N. Gehrels, \& B. Dennis
(Greenbelt: NASA Conf. Publ. no. 3137), p. 69} 

\jref{Harmon, B.A., \etal,  1995} Nature 374 {703 (H95)}

\jref{Hjellming, R.M. \& Rupen, M.P., 1995} Nature 375 {464 (HR95)}

\rref{Hjellming, R.M. and Rupen, M.P., 1996, IAU Circular no. 6411}

\rref{Horne, K., \etal, 1996,
IAU Circ. no. 6406}

\rref{Hunstead, R. \& Campbell-Wilson, D., 1996, IAU Circular no. 6410}

\jref{K\"onigl, A., 1981} Ap.J. 243 700

\jref{K\"onigl, A., 1989} Ap.J. 342 208
 
\rref{Inoue, H. \etal, 1994, IAU Circ. no. 6063}

\jref{Margon, B., 1984} Ann.Rev.Astron.\&Astrophys. 22 507
 
\rref{McKay, D. \& Kesteven, M., 1994, IAU Circ. no. 6062,
Aug. 26, 1994}
 
\jref{Mirabel, I.F., \& Rodriguez, L.F., 1994} Nature 371  46

\rref{Pacholczyk, A.G., 1969, Radio Astrophysics 
(San Francisco, Freeman)}

\rref{Remillard, R., Bradt, H., Cui, W., Levine, A., Morgan, E., Shirey, B. and
   Smith, D., 1996, IAU Circular no. 6393}

\jref{Sams, B.J., Eckart, A. \& Sunyaev, R., 1996} Nature 382 47

\jref{Tingay, S.J., \etal,  1995} Nature 374 {141 (T95)}

\rref{Wilson, C.A., Harmon, B.A.,  Zhang, S.N.,  
Paciesas, W.S. \&  
 Fishman  G.J., 1995, IAU Circ. no. 6152}
 
\rref{Zhang, S.N., Harmon, A.B., \etal, 1994, IAU Circ. no. 6046}

\rref{Zhang, S.N., \etal, 1996, to be submitted to ApJ}

\newpage

\cen{\bf Table 1}
\vskip .05in
\begin{center}
\begin{tabular}{|cccccc|}
\hline
$d$   & $f_o/f_c$ & $R_c$ & $R'_c$ &  $f(R'_c)$ & $L(R'_c)$ \\
(1)   & (2)       & (3)   & (4)      &  (5)     & (6)\\
\hline
$0\fs2$ & 0.03 & 19.9 & 16.6 &  $7.4\cdot 10^{-13}$ & $1.1\cdot 10^{33}$  \\
$0\fs5$ & 0.002 &22.8 & 19.5 & $5.1\cdot 10^{-14}$ & $7.4\cdot 10^{31}$  \\
$0\fs7$ & 0.0005&24.3 & 21.1 & $1.3\cdot 10^{-14}$ & $1.9\cdot 10^{31}$  \\
\hline
\end{tabular}
\end{center}

\vskip .3in
\ce{Table Caption}
\nn
April 1995 HST observations of \grop.
(1) angular distance from the central source (in arcseconds);
(2) ratio of the $R_c$-band  energy flux upper limit at the distance $d$
 to the flux from the central source;
(3) apparent $R_c$ magnitude upper limits of plasmoids
at the distance $d$ (not corrected for extinction);
(4) true $R_c$ magnitude upper limits of plasmoids 
corrected for extinction;
(5) upper limit to the energy flux 
(in $\rm erg \, cm^{-2} \, s^{-1}$) corresponding to the
$R'_c$ magnitude;
(6) upper limit to the  optical luminosity  
(in $\rm erg \, s^{-1}$) corresponding to the
$R'_c$ magnitude for a source at 3.5~kpc;

\newpage

\vskip 3.in

\ce{Figure Captions}

\vskip .2in

Fig. 1 - 
Combined  one-day averaged
 BATSE (20-100 keV)) (upper plot) and VLA 
(lower plot)
 lightcurves of \gro covering the time period August 1994-August 1995.
VLA data points (1.49 GHz) marked with filled black circles
and upper limits (4.9/8.4 GHz) marked with open triangles.
The March-April 1995 hard X-ray outburst is clearly evident, together with the
absence of detectable radio emission since late December 1994.
The dates of the HST observations of \gro (April 25 and 27 1995) 
are marked as downward pointing arrows in both plots.

\vskip .2in

\vskip .1in
Fig. 2 - 
HST-WFPC2 image (F675W filter) of \gro and of its surrounding field
obtained on April 25, 1995. \gro is the brightest source
slightly displaced from the center of the image.
Only the central source of apparent $R_c$ magnitude 
16.2 is detected.

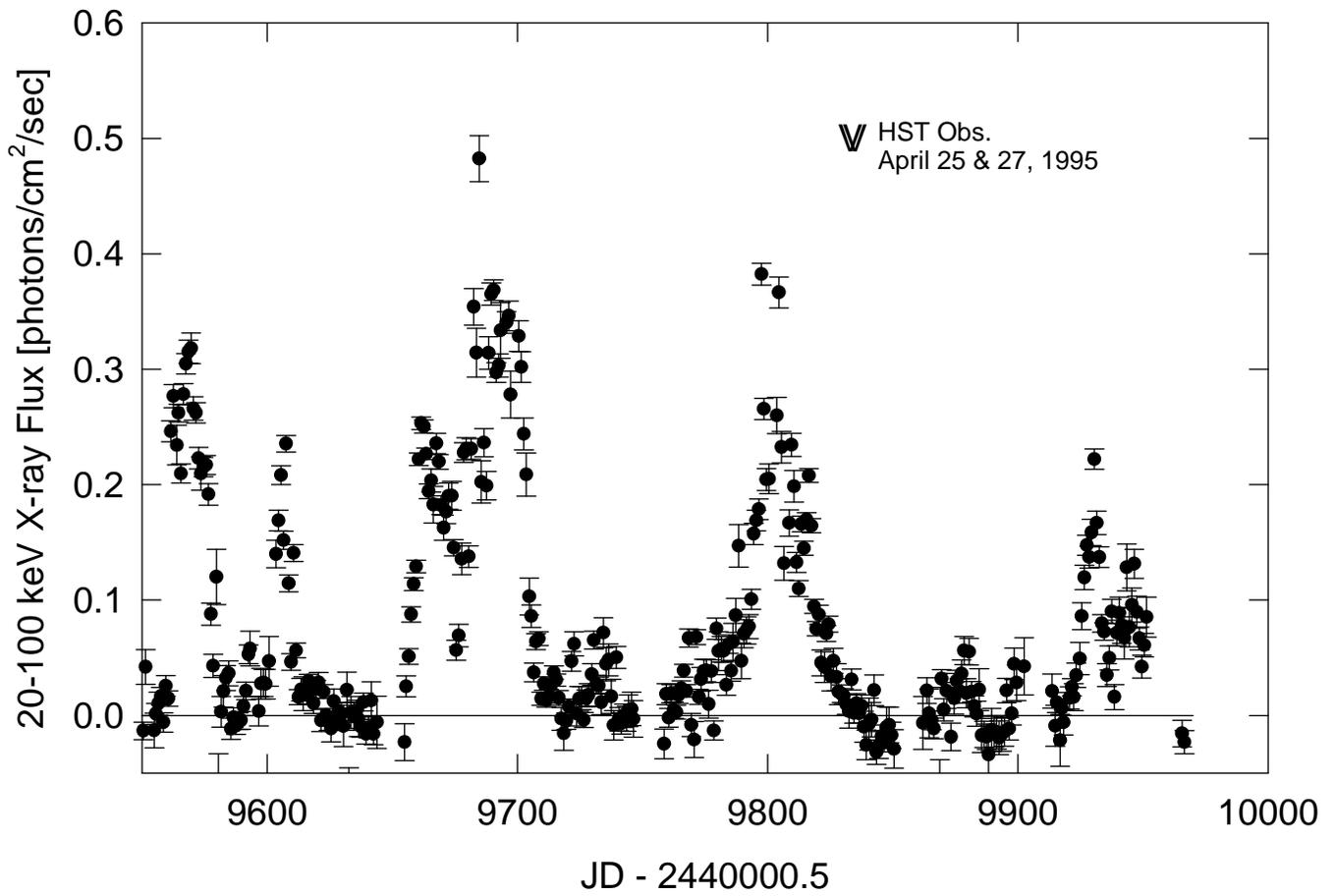
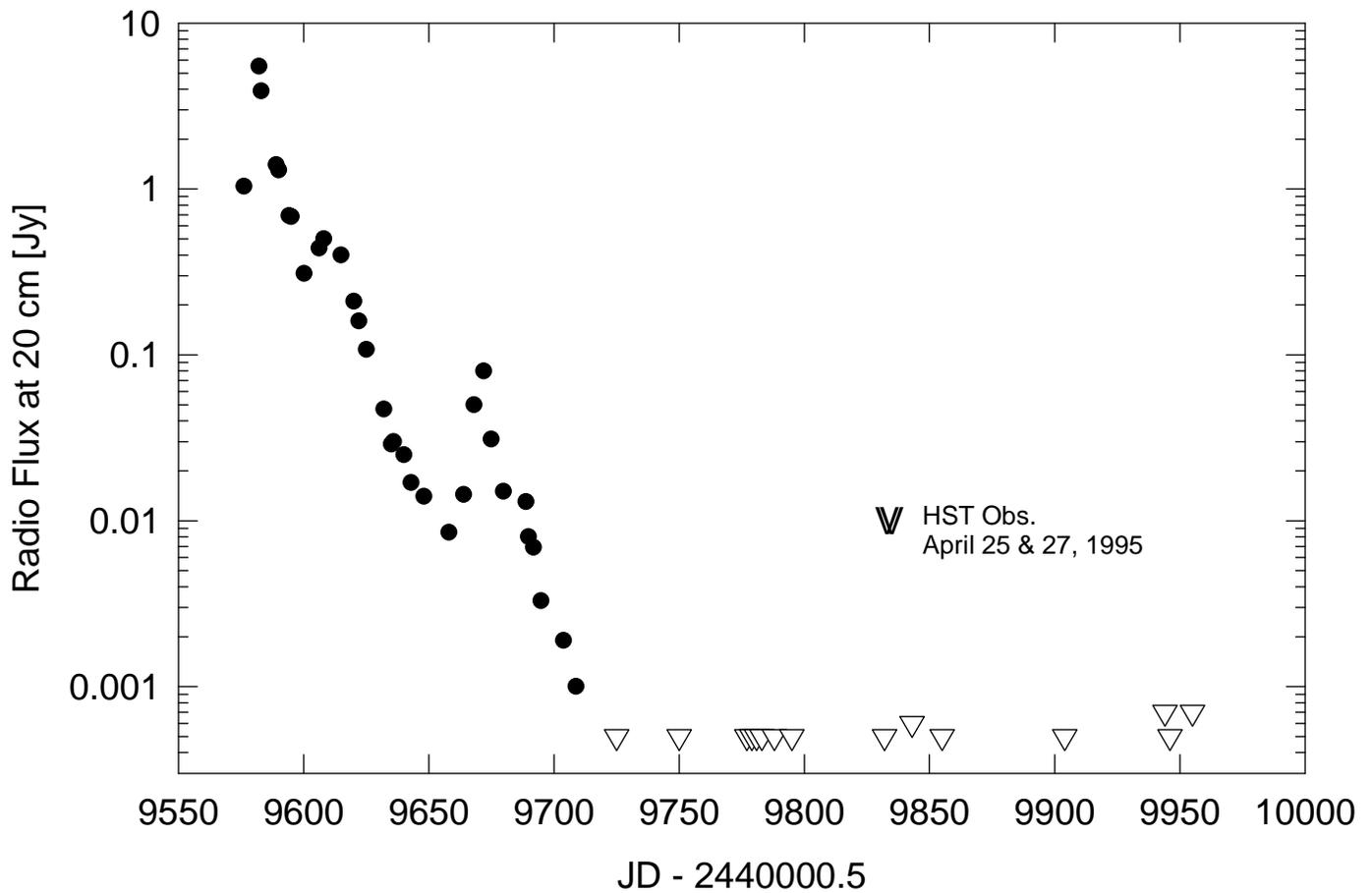
\end{document}